\RequirePackage{fix-cm}
\documentclass[twocolumn,compsoc]{CVM}
\usepackage{graphics}

\usepackage{float}
\usepackage{listings}
\usepackage{gensymb}

\usepackage{enumitem, hyperref}
\makeatletter
\def\namedlabel#1#2{\begingroup
	#2%
	\def\@currentlabel{#2}%
	\phantomsection\label{#1}\endgroup
}
\makeatother
 
\newcommand{\rev}[1]{{#1}}

\def \appendices{\par
    \gdef\theHsection{Appendix.\Alph{section}}
    \setcounter{section}{-1}
    \setcounter{subsection}{0}%
    \setcounter{paragraph}{0}%
    \gdef\thesection{\Roman{section}}%
    \gdef\thesectiondis{\Roman{section}}%
    \refstepcounter{section}
    \setcounter{section}{0}
}

\def\ManuscriptInfo{Manuscript received: 2014-12-31; accepted: 2015-01-30.}

\def\PaperTitle{Angle-Uniform Parallel Coordinates}

\def\AuthorNames{Kaiyi Zhang$^{1,*}$, Liang Zhou$^{2,*}$\textbf{(\rm{\Letter}\textbf{)}}, Lu Chen$^1$, Shitong He$^1$, Daniel Weiskopf$^3$,  Yunhai Wang$^1$}

\def\AuthorAdress{
    $\ast\quad $ & K. Zhang and L. Zhou are joint first authors. L. Zhou is the corresponding author.\\
    $1\quad $ & School of Computer Science and Technology, Shandong University, Qingdao 266237, China. \{zhang.kaiyi42, chenlu.scien,  heshitong2021, cloudseawang\}@gmail.com.\\
    $2\quad $ & National Institute of Health Data Science, Peking University, Beijing 100191, China. zhoul@bjmu.edu.cn.\\
    $3\quad $ & Visualization Research Center (VISUS), University of Stuttgart, 70569 Stuttgart, Germany. weiskopf @visus.uni-stuttgart.de.\\
}

\def\firstheader{DOI 10.1007/s41095-xxx-xxxx-x\hfill Vol. x, No. x, month year, xx--xx\\[5mm]~Research Article}

\headevenname{{K. Zhang, L. Zhou, L. Chen \etal} }

\definecolor{revBlue}{RGB}{77,139,183}

\begin{document}

\MakePageStyle

\MakeAbstract{
	We present angle-uniform parallel coordinates, a data-independent technique that deforms the image plane of parallel coordinates so that the angles of 
	\rev{linear relationships between two variables}
	are linearly mapped along the horizontal axis of the parallel coordinates plot.
	Despite being a common method for visualizing multidimensional data, parallel coordinates are ineffective for revealing positive correlations since the
	\rev{associated parallel coordinates points of such structures} may be located at infinity in the image plane and the asymmetric encoding of negative and positive correlations may lead to unreliable estimations.
	To address this issue, we introduce a transformation that bounds all points horizontally using an angle-uniform mapping and shrinks them vertically in a structure-preserving fashion; polygonal lines become smooth curves and a symmetric representation of data correlations is achieved.
	We further propose a combined subsampling and density visualization approach to reduce visual clutter caused by overdrawing.
	Our method enables accurate visual pattern interpretation of data correlations, and its data-independent nature makes it applicable to all multidimensional datasets.
	The usefulness of our method is demonstrated using examples of synthetic and real-world datasets.
}

\MakeKeywords{Parallel coordinates, multidimensional data, deformation, correlations}


\section{Introduction}\label{sec:introduction}

	\begin{figure}[t!]
	\centering
	\includegraphics[width=\linewidth]{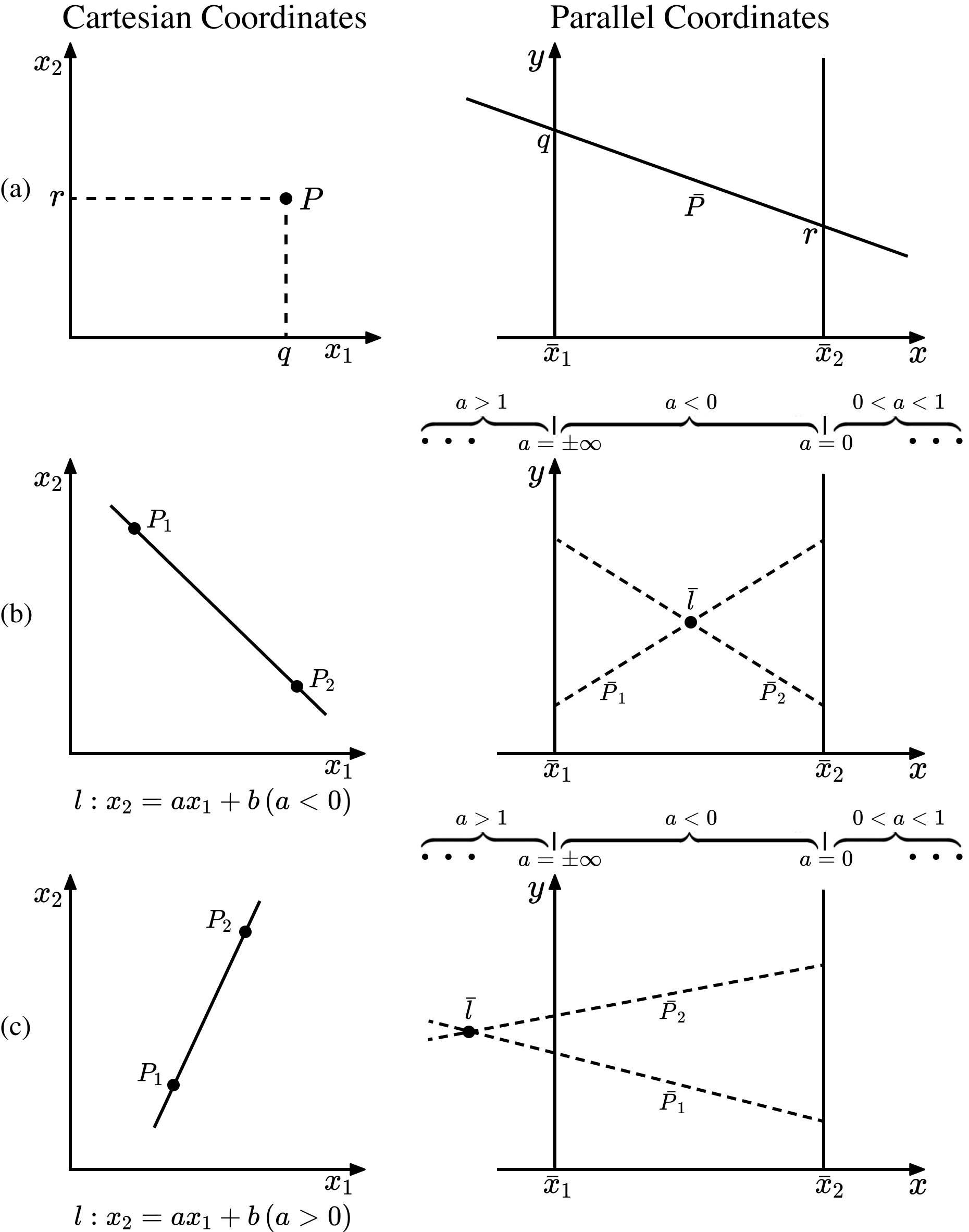}
	\caption{Point--line duality. (a) A point in Cartesian coordinates (i.e., a data point in a scatterplot) becomes a line in parallel coordinates, whereas (b) a line in Cartesian coordinates becomes a point in parallel coordinates: it is the intersection of a cluster of lines in parallel coordinates that represent a set of points on a Cartesian line. Points that represent Cartesian lines with negative slopes (negative correlations) lie in between their own pair of axes, whereas (c) points that represent positive correlations lie outside.}
	\label{fig:pointLineDual}
\end{figure}

Parallel coordinates are a popular visualization method for multidimensional data~\cite{Inselberg:1985:VisualComp,Wegman:1990:HDA}.
A data point in $n$-dimensional space is mapped to a polygonal line (polyline) with vertices on $n$ vertical axes, while a line in Cartesian coordinates is represented as an intersection point of polylines in parallel coordinates (see Fig.~\ref{fig:pointLineDual}).
The main benefit of parallel coordinates is their scalability in dimensionality: more axes just need to be added with increasing data dimensionality. A further advantage is that it is easy to trace data across multiple dimensions in parallel coordinates.

However, a key issue when using parallel coordinates is the difficulty of visual pattern interpretation, especially for positive correlations. 
\rev{
	In this paper, we refer to the 2D space of a scatterplot as Cartesian space, in which Cartesian coordinates represent the two data attributes describing a data point. An ideal linear correlation shows up as a straight line in a scatterplot; we refer to this as a Cartesian line (see Fig.~\ref{fig:pointLineDual}(b) and (c), left column).
	In parallel coordinates, the associated points of Cartesian lines with positive slopes lie outside of their own pair of axes (see Fig.~\ref{fig:pointLineDual}(c)).}
Furthermore, if the angle of a line is $45\degree$, the corresponding point would be at infinity in the image plane,  making it impossible for users to observe.

We further note that linear and symmetric encodings are preferred by humans for quantitative analysis.
However, the mapping between the angles of Cartesian lines and the horizontal coordinates of associated points in parallel coordinates given by the point--line duality (Section~\ref{sec:pointlineDual}) is nonlinear and asymmetric.
As a result, Li et al.~\cite{Li:IV:2010} reported a perception bias toward negative correlations and a general underestimation of both negative and positive correlations of users of parallel coordinates, as shown by a user experiment.

Overdraw is another issue when using parallel coordinates.
Visual patterns are clear and traceable in parallel coordinates when the scale of the data is small.
However, the visualization quickly becomes cluttered with larger datasets as visual patterns are obscured by the overlapping polylines.
These drawbacks greatly hamper the readability of parallel coordinates and limit their applicability.

There are techniques to visualize  local linear relationships in parallel coordinates~\cite{Nguyen:TVCG:2017,Zhou:TVCG:inprint}.
Unfortunately, the asymmetric representation of Cartesian lines is not fully addressed by either technique, as different visual mappings are used for negative and positive relationships~\cite{Nguyen:TVCG:2017} and positive linear correlations with slopes close to 1 still cannot be visualized in the same way as intersection points showing negative correlations~\cite{Zhou:TVCG:inprint}.
We thus present angle-uniform parallel coordinates, a novel method that achieves a symmetric and unified representation of data correlations in a confined image plane.

\begin{figure*}[t!]
	\centering
	\includegraphics[width=\linewidth]{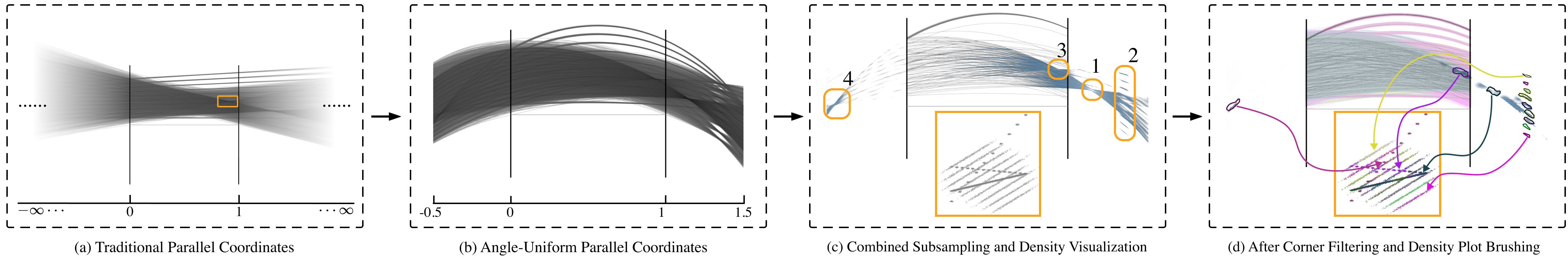}
	\caption{\rev{Processing pipeline of angle-uniform parallel coordinates, using a synthetic dataset containing several linearly correlated structures (see Section~\ref{sect:synthetic_data} for details of the data). (a)~Traditional parallel coordinates have an image plane that extends to infinity. (b)~Angle-uniform parallel coordinates deform the image plane into a bounded, structure-preserving 2D space, providing a symmetric representation of negative and positive correlations. (c)~The combined subsampling and density visualization reduces visual clutter while preserving global and local patterns. Patterns indicating strong correlations are marked with numbered rounded boxes; the orange square below the parallel coordinates show a scatterplot of the data for reference. (d)~Interactive exploration of the new visualization is aided by corner filtering and density plot brushing as demonstrated by the associations between brushed regions and their corresponding structures in the scatterplot. 
	}}
	\label{fig:synthetic}
\end{figure*}

The first contribution of this paper is the angle-uniform parallel coordinates model \rev{(see Fig.~\ref{fig:synthetic}(b))}.
By transforming the image plane of parallel coordinates in two directions, this model maps parallel polylines to smooth, intersecting curves.
Specifically, all points are deformed into a bounded area by a linear mapping of angles in the horizontal direction.
In the vertical direction, the points are shrunk in a structure-preserving fashion to ensure the smoothness and symmetry of the deformed curves.
Furthermore, no information from the traditional parallel coordinates is lost during the transformation, and the relative relationships between original data points and the continuity across different dimensions are carefully preserved.

Our second contribution is a combined subsampling and density visualization approach \rev{(see Fig.~\ref{fig:synthetic}(c))} that reduces clutter to facilitate correlation visualization in angle-uniform parallel coordinates.
The combination of subsampled data curves and density plots allows a coherent and clear visualization of both significant global patterns and important local patterns.

\rev{Our third contribution is an interactive system that supports corner filtering and density plot brushing (Fig.~\ref{fig:synthetic}d), two new interaction techniques specially designed for combined subsampling and density visualization, to facilitate exploratory data analysis.}


Using a synthetic dataset, we \rev{compare different variants of parallel coordinates and} show that users can produce accurate interpretations of negative and positive correlations alike using angle-uniform parallel coordinates.
To further demonstrate the usefulness of our method, we provide further visual analysis examples of real-world multidimensional datasets, using traditional parallel coordinates, scatterplots, and angle-uniform parallel coordinates.
Compared to other parallel coordinates methods, our angle-uniform parallel coordinates are data-independent, so they can be applied to any multidimensional dataset and easily incorporated within existing parallel coordinate systems.

\section{Related Work}\label{sec:RelatedWork}
Parallel coordinates are commonly used for visualizing multidimensional data.
Early works advocating such methods can be traced back to Inselberg~\cite{Inselberg:1985:VisualComp} and Wegman~\cite{Wegman:1990:HDA}.
Inselberg formulates comprehensive mathematical descriptions of geometric entities in parallel coordinates and uses them to visualize multidimensional geometry~\cite{InselbergPCPbook}. 
Wegman applies parallel coordinates to visual data mining of multidimensional data~\cite{Wegman:1990:HDA}.
Both works study parallel coordinates from the perspective of projective geometry and represent parallel coordinates using homogeneous coordinates~\cite{InselbergPCPbook,Wegman:WICS122}.
A survey of the state of the art of parallel coordinates can be found in~\cite{HeinrichPCPEG13}.

\subsection{Correlation Analysis with Parallel Coordinates}
The basic and most popular visualization approach using parallel coordinates is to display only polylines that correspond to multidimensional data points.
Correlations are shown as high-density intersections of polylines, which are, however, asymmetric for negative and positive correlations. 
The asymmetry leads to the underestimation of positive correlations by users~\cite{Li:IV:2010}.
Hybrid methods that combine parallel coordinates with other visual mappings, such as scatterplots, scatterplot matrices (SPLOMs), or variants thereof, improve the ability to find and analyze correlations. 
Scatterplots can be drawn next to parallel coordinates to show pair-wise correlations of data attributes~\cite{Qu:VIS2007,Steed:VAST09,Holten:CGF:CGF1666}.
Alternatively, scatterplots can be embedded between adjacent axes in parallel coordinates~\cite{Yuan:VIS:2009}.
The P-SPLOM technique~\cite{Viau:TVCG:2010} unifies SPLOM, 2D parallel coordinates, and 3D parallel coordinates with smooth transitions between them. 
Flexible user-customized visualizations in the style of parallel coordinates or SPLOMs are available~\cite{Claessen:TVCG:2011}.

Local linear relationships can be directly visualized in parallel coordinates. 
One technique~\cite{Nguyen:TVCG:2017} estimates local linear relationships to visualize correlations in parallel coordinates using the point--line duality. 
Another method~\cite{Zhou:TVCG:inprint} uses $p$-flat indexed points, compact representations of $p$-dimensional generalized flat surfaces, to visualize local multidimensional linear relationships in parallel coordinates.
While negative correlations are shown effectively, visual mappings of positive correlations have their own limitations in both techniques. 
Furthermore, both works provide a nonlinear relationship between the slopes of Cartesian lines and the positions of intersections in parallel coordinates.

\subsection{Curved Parallel Coordinates}
Our approach leads to a deformation of the plane of parallel coordinates and, thus, transforms polylines to curves. 
Curves have been used before to replace polylines, but for different reasons and using different curve models to ours. For example, curves can be designed to facilitate easier tracing than polylines~\cite{Graham:iv:2003,Yuan:VIS:2009,Holten:CGF:CGF1666}.
As another example, curve representations with bundling enable the visualization of clusters~\cite{Zhou:eurovis:2008, McDonnell:eurovis:2008,Heinrich:2012:EBT}, 
and the visualization of multiple and higher-order correlations can be aided by the use of curves~\cite{Theisel:2000:HOP}.
However, important geometric information in polyline parallel coordinates is lost in the above works: for example, intersection patterns are lost. 
Unlike existing curved parallel coordinates, our technique preserves important geometric properties from the original polyline-based parallel coordinates.

The transformation from traditional parallel coordinates to our angle-uniform parallel coordinates is closely related to hyperbolic geometry.
Hyperbolic geometry~\cite{anderson2005hyperbolic} is a non-Euclidean geometry that satisfies all of Euclid's postulates except for the parallel postulate.
In hyperbolic geometry, parallel lines pass through a common point. Therefore, an infinite space in Euclidean geometry can be represented in a limited hyperbolic space.
A very well-known example of the usage of hyperbolic space is the painting series `Circle Limit' by Escher~\cite{escher:website}.
In the context of visualization, hyperbolic geometry has been used for visualizing large graphs with the focus+context approach~\cite{Munzner:infovis:1997}.
In contrast, our hyperbolic-inspired mapping generates a static space that introduces no context change, and therefore preserves the mental map of users.

\subsection{Clutter Reducing Visualization}
Parallel coordinates quickly become cluttered with increasing number of data points.
Subsampling from the full dataset and density representations are two general approaches to reducing clutter in visualization~\cite{Ellis:TVCG07}.
By subsampling, local patterns can be seen in parallel coordinates, but it cannot faithfully convey  global information.
The density approach replaces line plots by density representations, e.g., by different kinds of binning methods~\cite{Artero:2004:UCC,Johansson:2005:Infovis,Novotny:2006:Infovis} or continuous modeling~\cite{Heinrich:VIS:2009}.
However, this can blur the visualization as a whole and remove interesting local patterns, making correlation recognition harder in parallel coordinates.
Our new combined subsampling and density visualization integrates the benefits of curve representation of subsampled data and the advantages of density representation.


\section{Background and Rationale}\label{sec:theory}
In this section, we explain the theoretical foundation of parallel coordinates, namely point--line duality, and the rationale of our method. 
More details of geometric analysis in parallel coordinates can be found in the book by Inselberg~\cite{InselbergPCPbook}.

\subsection{Point--Line Duality}
\label{sec:pointlineDual}
The point--line duality specifies the mapping relation of basic geometry entities, i.e., points and lines, from Cartesian coordinates to parallel coordinates.
In parallel coordinates, data points in $n$ dimensions are usually represented as polylines with vertices on $n$ consecutively placed axes in the 2D image plane.
To create a parallel coordinates plot, an $n$-dimensional dataset is split into $n-1$ independent 2D subspaces, and each subspace is then plotted in the image plane defined by two adjacent parallel axes.

We introduce the point--line duality using an elementary case of a 2D subspace spanned by $x_1$ and $x_2$.
Here, we denote the horizontal and vertical coordinates of a point in the parallel coordinates image plane  as $x$ and $y$, respectively.
We assume that the origin of the image plane is at $x_1 = 0$, and the distance between the two parallel axes ($\bar{x}_1$ and $\bar{x}_2$) is $1$. 
In Cartesian coordinates, a 2D data item $(q,r)$ is represented as a point $P$.
The representation of point $P$ in parallel coordinates is a line segment $\bar{P}$ with  endpoints $(0,q)$ and $(1,r)$ (see Fig.~\ref{fig:pointLineDual}(a)).

In contrast, a line in Cartesian coordinates is mapped to a point in parallel coordinates.
In Cartesian coordinates, a line $l$ can be described as:
\begin{align}
	l: x_2 = ax_{1} + b,
\end{align} 
where $a$ and $b$ are the slope and intercept of $l$, respectively.
To derive the representation of $l$ in parallel coordinates, we pick two points $P_1$ and $P_2$ from $l$ and calculate their corresponding line segments $\bar{P}_1$ and $\bar{P}_2$, and find the intersection point $\bar{l}$ as shown in Fig.~\ref{fig:pointLineDual}(b, c).
Therefore, point $\bar{l}$ is the parallel coordinates representation of line $l$ in Cartesian coordinates~\cite{InselbergPCPbook}:
\begin{align}
	\label{eqn:lineInPCP}
	\bar{l} = \left(x, y\right) = \left(\frac{1}{1-a},\quad \frac{b}{1-a}\right), a \neq 1.
\end{align}

These point-to-line and line-to-point mappings comprise the point--line duality, which is a fundamental property of parallel coordinates.
In fact, the point representation $\bar{l}$ is also called the \emph{ 1-flat indexed point} of $\bar{l}$~\cite{InselbergPCPbook,Zhou:TVCG:inprint}, and this terminology is adopted in our paper. 

To facilitate the discussion, we write $l$ in its general implicit form to achieve complete generality: 
\begin{equation}
	l: c_1x_1 + c_2x_2 + c_3 = 0\;.
\end{equation}
We can express $l$ by its parameters as \emph{line coordinates}, i.e., a tuple $[c_1,c_2,c_3]$ in homogeneous coordinates.
The associated indexed point $\bar{l}$ can also be described in parallel coordinates by a tuple in homogeneous coordinates.
Therefore, the following relationships between the triples $l$ and $\bar{l}$ can be obtained: 
\begin{equation}
	l:[c_1, c_2, c_3] \to \bar{l}: 
	\begin{cases}
		(c_2, -c_3, c_1+c_2), &c_2\neq0\;,\\
		(0, -{c_3}/{c_1}, 1), &c_2 = 0\;.
	\end{cases}
	\label{eqn:homoPCP}
\end{equation}   

\begin{figure}[t!]
	\centering
	\includegraphics[width =\linewidth]{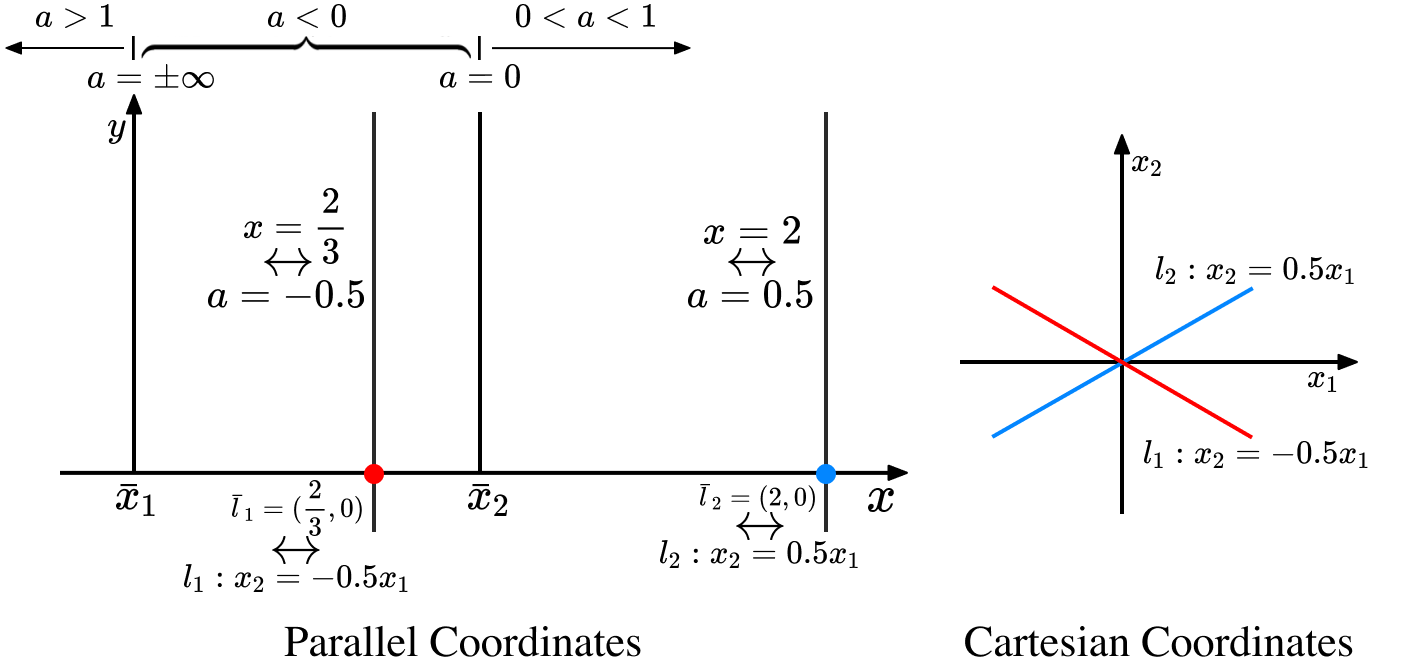}
	\caption{Asymmetric representation of 1-flat indexed points with negative and positive slopes.}
	\label{fig:asymexample}
\end{figure}

\subsection{Mapping Rationale}
\label{sec:mapCriteria}
The point--line duality results in an asymmetric representation of 1-flat indexed points in parallel coordinates associated with negative and positive correlations in Cartesian coordinates.
For example, as shown in Fig.~\ref{fig:asymexample}, the indexed point $\bar{l}_1$ of line $l_1: x_2 = -0.5x_1$ is at $({2}/{3},0)$, whereas the indexed point $\bar{l}_2$ of line $l_2: x_2 = 0.5x_1$ is at $(2,0)$, and they are not symmetric about the $\bar{x}_2$ axis.
Vertical lines $x={2}/{3}$ and $x=2$ indicating all Cartesian lines with slopes $a=-0.5$ and $a=0.5$ are also drawn in Fig.~\ref{fig:asymexample}.
As the slope of a line approaches +1, the location of its indexed point approaches positive or negative infinity.
Therefore, interpretation of line information from the location of an indexed point of a line with a positive slope close to +1 is difficult.

We address the asymmetry issue with a nonlinear mapping.
We would like all parallel lines except for vertical ones to have intersections in our deformed space. 
Vertical lines are kept non-intersecting by design to be consistent with the behavior of attribute axes in traditional parallel coordinates. 

\begin{figure*}[!t]
	\centering
	\includegraphics*[width=\linewidth]{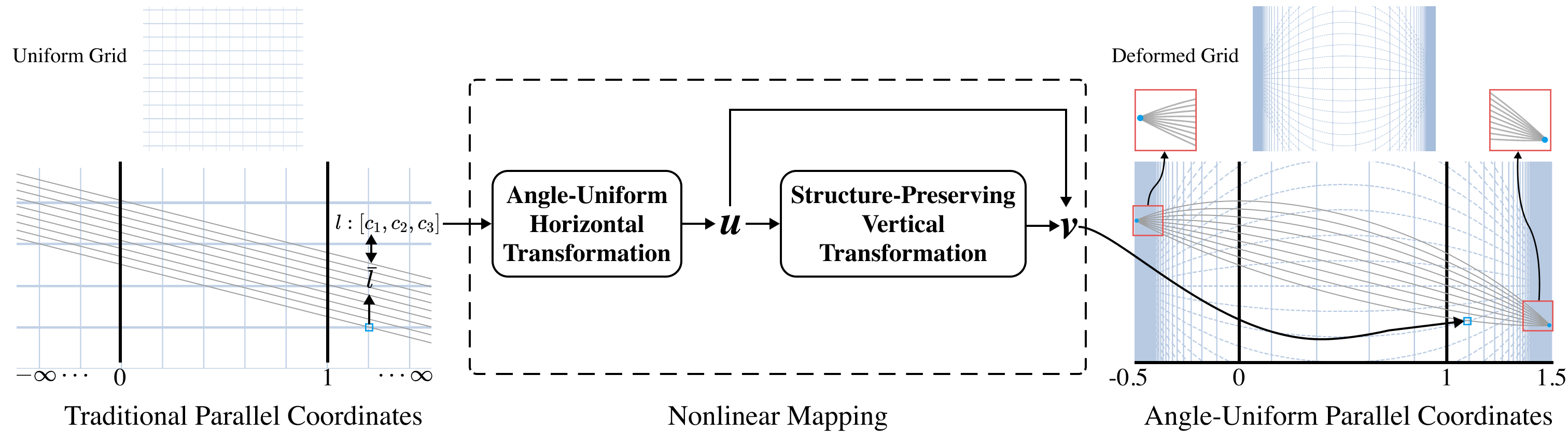}
	\caption{Transformation pipeline: infinitely long parallel lines in traditional parallel coordinates are mapped to curved lines of limited length that intersect each other in angle-uniform parallel coordinates. The transformation is applied to every point in the 2D image plane of traditional parallel coordinates: an example is illustrated by a point $\bar{l}$ (highlighted in a blue box). The underlying uniform grid of traditional parallel coordinates is deformed to a curvilinear grid that is bounded horizontally in angle-uniform parallel coordinates (blue, at top). }
	\label{fig:abalone_synth}
\end{figure*}

Using our approach, all 1-flat indexed points, including those of slope +1, can be displayed in a limited 2D area. 
Fig.~\ref{fig:abalone_synth} shows a synthetic dataset with two identical attributes, resulting in a perfectly correlated line of slope +1 in Cartesian coordinates. 
This leads to parallel lines in traditional parallel coordinates (see Fig.~\ref{fig:abalone_synth}(left)), where no indexed points can be seen because they are at positive or negative infinity. 
After our nonlinear mapping (see Fig.~\ref{fig:abalone_synth}(right)), parallel lines become curves that intersect at horizontal locations at -0.5 and 1.5 in the 2D image plane, and all 1-flat indexed points (shown in blue) are located at these two intersection points, indicating a perfect correlation.   
Fig.~\ref{fig:abalone_synth}(top) shows how the underlying grid of the 2D image plane of parallel coordinates is deformed.
Furthermore, the orientation of the line associated with a 1-flat indexed point can now be read off and converted to angles (as shown in Fig.~\ref{fig:slopenOrientation}(b)) from the visualization as shown in the zoomed-in insets of Fig.~\ref{fig:abalone_synth}.

\section{Method}\label{sec:method}
In this section, we derive our transformations of parallel coordinates step by step.
For the ease of description, we start by explaining the transformation of 1-flat indexed points in parallel coordinates, and extend the transformation to lines afterwards.
\rev{The transformation pipeline is shown in Fig.~\ref{fig:abalone_synth}.}

\subsection{Design Considerations}
Ideally, the new parallel coordinates model should offer symmetric patterns of positive and negative correlations while preserving the benefits of traditional parallel coordinates: data traceability across dimensions, and dimensionality scalability.
To achieve these, we list the design goals of our method as follows:
\begin{itemize}[noitemsep, topsep=3pt]
	\item[\namedlabel{itm:infty}{\textbf{G.1}}] The infinite horizontal range of parallel coordinates should be bounded.
	\item[\namedlabel{itm:linear}{\textbf{G.2}}] Orientations of Cartesian lines should map linearly to horizontal positions.
	\item[\namedlabel{itm:vert}{\textbf{G.3}}] The relative relationship between  vertical and horizontal coordinates in the original parallel coordinates should be preserved.
	\item[\namedlabel{itm:subspaceC}{\textbf{G.4}}] A curve transformed from a line segment in a subspace of parallel coordinates should have $C^1$ continuity.
	\item[\namedlabel{itm:crossspaceC}{\textbf{G.5}}] Curves transformed from a polyline that represents a multidimensional data sample should have $C^0$ continuity across different attribute axes.
	\item[\namedlabel{itm:symmetric}{\textbf{G.6}}] 
	\rev{The curve transformed from a line segment in a subspace should be symmetric under exchange of adjacent attribute axes.}
\end{itemize}

\begin{figure}[htb]
	\centering
	\includegraphics[width = \linewidth]{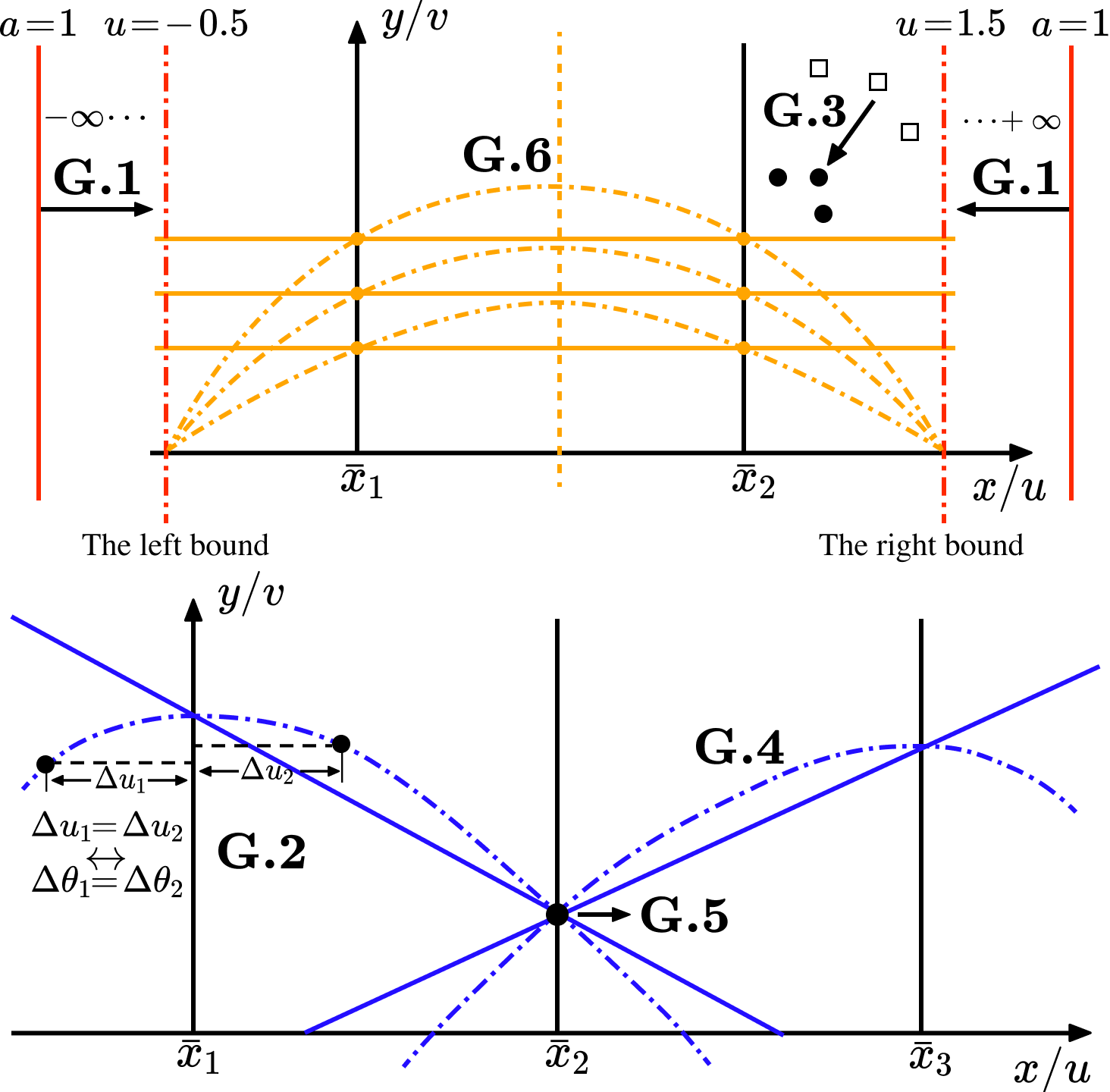}
	\caption{Design goals \textbf{G.1}--\textbf{G.6} of our transformation. Solid colored lines indicate lines in the traditional parallel coordinates, dashed lines indicate curves in the transformed parallel coordinates. Hollow squares and filled disks indicate 1-flat indexed points of the traditional and transformed parallel coordinates, respectively.}
	\label{fig:xfgoals}
\end{figure}

The meanings of these goals are illustrated in Fig.~\ref{fig:xfgoals}.
\textbf{G.1} is the basic goal of bounding parallel coordinates horizontally.
\textbf{G.2} allows direct and intuitive perception of orientations of Cartesian lines in the deformed space.
Patterns in the original parallel coordinates are preserved by~\textbf{G.3}. 
To trace a data item within the image plane formed by a pair of parallel axes, the data has to be drawn as a smooth curve as in~\textbf{G.4}.
\textbf{G.5} is necessary for tracing data items across multiple parallel axes when two curves with the same data values meet at the same vertical location.
The need for symmetric representations along with aesthetic considerations leads to~\textbf{G.6} that the 
\rev{curve should be symmetric under exchange of adjacent attribute axes, i.e., curves before and after the swap should be mirrored at the center of the attribute pair, and a horizontal curve before transformation should be mirror-symmetric at the center of the attribute pair.}

\subsection{Transformation of 1-flat Indexed Points}
We consider a 1-flat indexed point $\bar{l}$ of line $l: c_1x_1 + c_2x_2 + c_3 = 0$. 
Using the line coordinates of $l : [c_1, c_2, c_3]$, we denote the mapping by:
\begin{equation*}
	F: [c_1, c_2, c_3] \mapsto (u,v), \quad c_1,c_2,c_3 \in\mathbb{R}, \quad (u,v) \in \mathbb{R}^2,
\end{equation*}
where $(u,v)$ are the output point coordinates in the 2D image plane of parallel coordinates.

\subsubsection{Angle-Uniform Horizontal Transformation}
\label{sec:horXf}
\begin{figure}[t!]
	\centering
	\includegraphics[width = \linewidth]{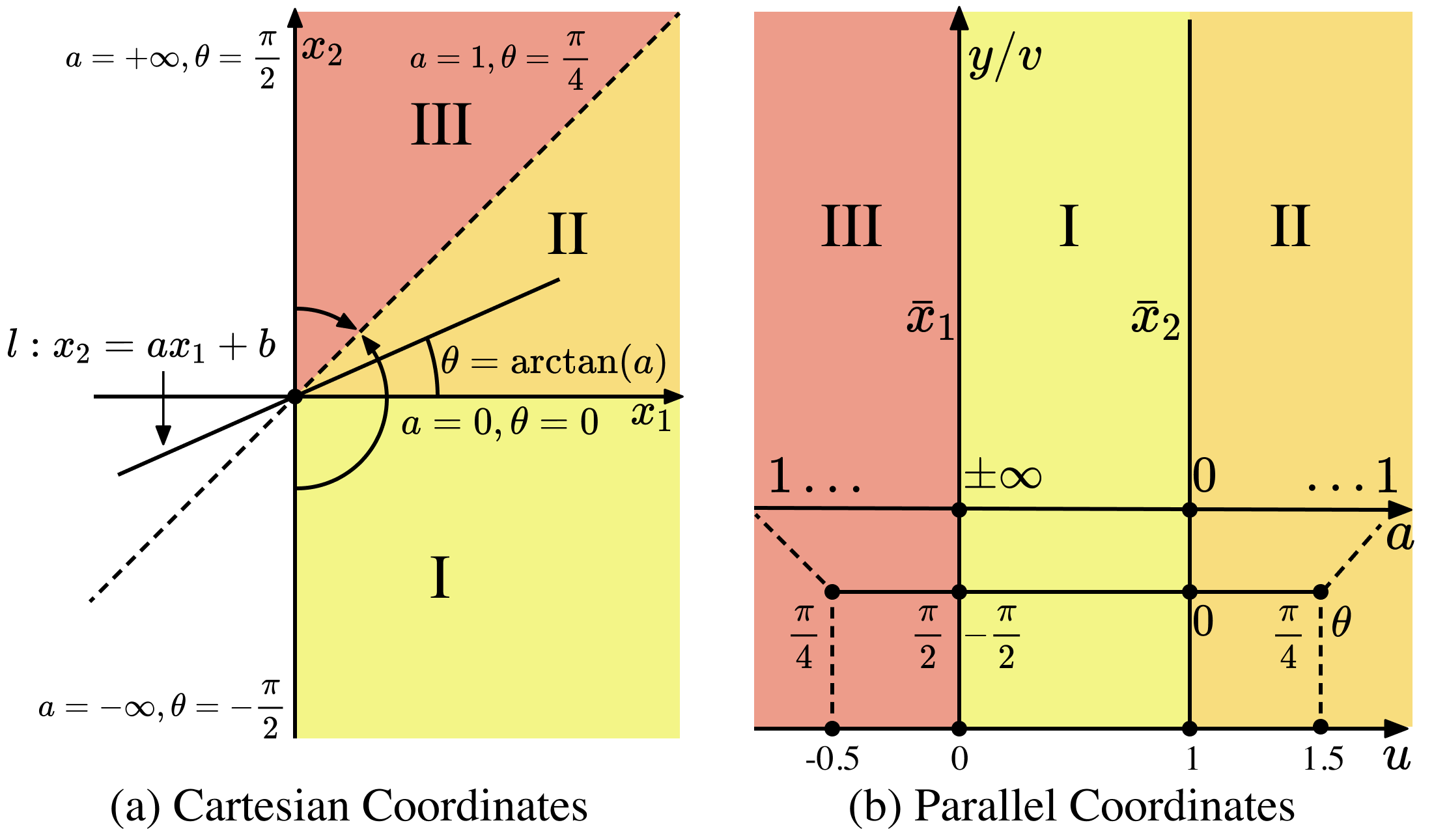}
	\caption{Relationship between the slope $a$ and the orientation $\theta$ of a line $l$ is shown in Cartesian coordinates (a) and parallel coordinates (b). The 2D image planes are divided into three regions: I, II, and III. In traditional parallel coordinates, the horizontal extents of regions II and III are infinite. In contrast, in the orientation representation, regions II and III become bounded horizontally.}
	\label{fig:slopenOrientation}
\end{figure}

The goals of the horizontal transformation are to transform 1-flat indexed points into a finite horizontal range (\ref{itm:infty}), using a linear representation for the horizontal coordinate (\ref{itm:linear}).
We start by replacing the horizontal coordinate of parallel coordinates from a slope-based metric to an orientation-based one.
The relationship between the slope $a$ and the orientation $\theta$ of $l$ in Cartesian coordinates is illustrated in Fig.~\ref{fig:slopenOrientation}(a).

From the line coordinate description $[c_1, c_2, c_3]$, we may compute $\theta$ as:
\begin{equation}
	\theta = 
	\begin{cases}
		\pm{\pi}/{2}, &c_2=0,\\
		\arctan(-{c_1}/{c_2}), &c_2\neq 0.
	\end{cases}
	\nonumber
\end{equation}
Then, we must transform $\theta$ to the 2D image coordinate $u$ to match the distance between the pair of parallel axes.
This is done by scale-and-bias:
\begin{equation}
	\boxed{
		u = 
		\begin{cases}
			{2\theta}/{\pi} - 1,&\theta > {\pi}/{4},\\
			{2\theta}/{\pi} + 1,&\theta < {\pi}/{4},\\
			-0.5\; \text{and}\; 1.5,&\theta = {\pi}/{4}.\\
		\end{cases}
	}
	\label{eqn:horixf}
\end{equation}
We set the left bound of $u$ to -0.5 and the right bound to 1.5 to build a symmetric representation in the horizontal coordinate.
Fig.~\ref{fig:slopenOrientation}(b) illustrates the relationship between $a$, $\theta$, and $u$.
With this horizontal transformation, goals~\ref{itm:infty} and \ref{itm:linear} are achieved.


\subsubsection{Structure-Preserving Vertical Transformation}
\label{sec:vertXf}
A proper vertical transformation has to be designed to achieve the remaining design goals.
In traditional parallel coordinates (with 2D point coordinates described by Equation~\ref{eqn:lineInPCP}), the vertical coordinate is related to the horizontal coordinate by the intercept $b$ of the line for $a\neq1$:
\begin{equation*}
	b = {y}/{x},\quad x\neq 0.
\end{equation*}
\rev{To preserve the relative ratio between horizontal and vertical coordinates of traditional parallel coordinates, we define the transformed vertical coordinate $v$ as:
	\begin{equation}
		v = (u - 0.5) \frac{y}{(x - 0.5)}, \quad u\neq -0.5, 0.5, 1.5.
		\label{eqn:firstv}
	\end{equation} 
}
The origin of the transformation is set to the center of the axis pair ($x=u=0.5$) to meet the goal of symmetry (\ref{itm:symmetric}).
To fully define $v$ in the 2D image plane of parallel coordinates, we need to handle two special cases not covered by Equation~\ref{eqn:firstv}.
\rev{First, for $u=0.5$, i.e., 1-flat indexed points with $x = 0.5$, $a=-1$ is located at the center of the attribute pair, leaving $v$ undefined.}
Second, for indexed points at infinity ($a=1$), we have mapped the transformed horizontal coordinate to -0.5 and 1.5 in Section~\ref{sec:horXf}, but the vertical coordinate is not yet defined.
We address these exceptions using the limit method, i.e., taking the vertical coordinate of a transformed point whose horizontal coordinate is infinitely close to the point of question as the vertical coordinate. 

\begin{figure}[!htb]
	\centering	
	\includegraphics[width = 0.9\linewidth]{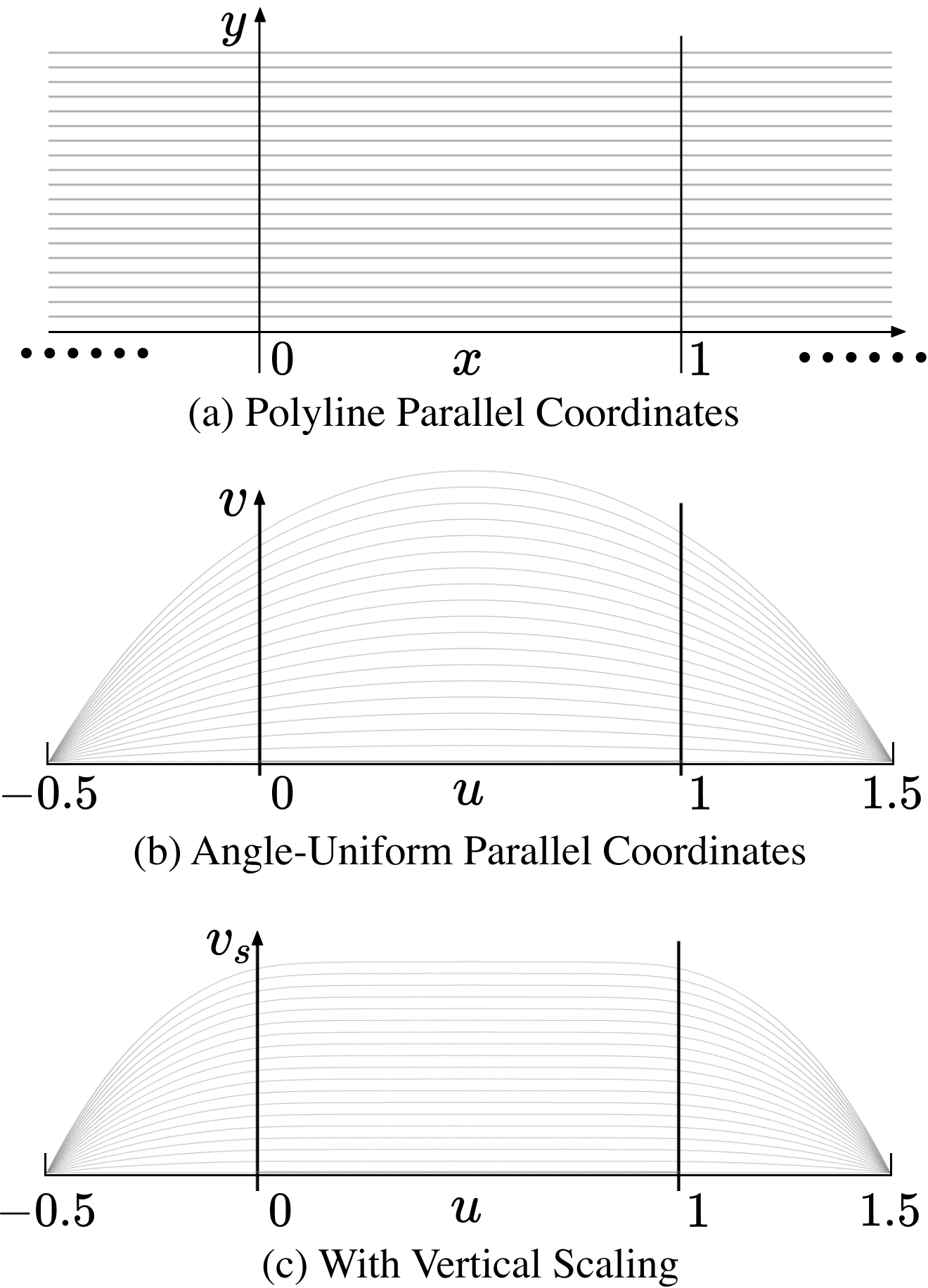}
	\caption{Vertical mapping without scaling when applied to a synthetic dataset containing identical attributes transforms the parallel lines in~(a) to curves~(b). Using the vertical scaling function, especially the curves within the attribute pair are flattened~(c).}
	\label{fig:vertMapSimple}
\end{figure}

Then, the transformed vertical coordinate considering all cases is defined as:
\begin{equation}
	\boxed{
		v = 
		\begin{cases}
			v_{-0.5}, &u = -0.5\\
			v_{0.5}, &u = 0.5\\
			v_{1.5}, &u = 1.5\\
			\frac{(u-0.5)y}{(x-0.5)}=\frac{2c_3(u-0.5)}{(c_1-c_2)}\;,&\text{otherwise}\;.\\
	\end{cases}}
	\label{eqn:fullv}
\end{equation}
A detailed derivation of Equation~\ref{eqn:fullv} is provided in Appendix~\ref{sec:apdvertXf}.
This mapping preserves the vertical versus horizontal information in traditional parallel coordinates as shown in Fig.~\ref{fig:vertMapSimple}. 
As seen in Fig.~\ref{fig:vertMapSimple}(b), transformed curves preserve the relative relationships of lines in traditional parallel coordinates (see Fig.~\ref{fig:vertMapSimple}(a)), satisfying~\ref{itm:vert} \rev{in the horizontal direction};~\ref{itm:subspaceC} is also achieved as transformed curves are smooth.
\rev{Moreover, the curves are mirror symmetric about the the center of the attribute pair, satisfying~\ref{itm:symmetric};~\ref{itm:crossspaceC} is achieved as transformed vertical values $v$ on parallel axes ($u=0, 1$) are equal to the original values $y$. }

\rev{In some cases, round curves yielded by the vertical transformation may need to be flattened within the attribute pair to retain the look of traditional parallel coordinates.
	Therefore, an optional scaling function $s(u)$ can be included as a factor in Equation~\ref{eqn:fullv}:}
\begin{equation}
	v_s = s(u) v.
\end{equation}
This scaling function is included to provide more flexibility in curve design to the user, but is not required.

\subsubsection{The Vertical Scaling Function}
\rev{We study how to flatten the vertical coordinate of Equation~\ref{eqn:fullv} by taking point samples on a horizontal line $y = C$ in parallel coordinates.}
The ratio $r$ between the transformed and original vertical coordinates can be written as:
\begin{equation}
	r = \frac{v}{y}  = \frac{(u-0.5)}{(x-0.5)},\quad -0.5\leq u \leq 1.5\;.
	\label{eqn:vvsy}
\end{equation}

We treat the three regions (I, II, III in Fig.~\ref{fig:slopenOrientation}) of the 2D image plane separately; a derivation of the piecewise scaling function $g(u)$ is provided in the appendix. 
If we multiply $v$ by $g(u)$, goals~\ref{itm:subspaceC} and~\ref{itm:crossspaceC} are not satisfied because the function is not continuous at $u=0$ and $u=1$.

This is an over-constrained problem that we resolve  by use of cubic spline~\cite{de1978practical} interpolation to ensure smoothness of the scaling function. 
\rev{The eleven control points that the derived cubic spline $s(u)$ must pass through are detailed in Appendix~\ref{sec:apdsu}.}
The cubic spline scaling function $s(u)$ has $C^2$ continuity and satisfies goals~\ref{itm:subspaceC}--\ref{itm:symmetric}.
The full transformation with the scaling function $s(u)$ applied to the synthetic dataset is shown in Fig.~\ref{fig:vertMapSimple}c.

Please note that most example images of this paper do not employ vertical scaling: the only exceptions are Fig.~\ref{fig:vertMapSimple}(c) and Fig.~\ref{fig:teaser}(b).

\subsection{Transformation of Parallel Coordinates Lines}
\label{sec:a1Xf}
A parallel coordinate line $\bar{P}$ can be transformed by applying point transformations to a set of samples on the line.
We sample the parallel coordinate line at equal angle intervals $\Delta \theta$.
We then obtain homogeneous coordinates $[c_1, c_2, c_3]$ for each sample as detailed in Appendix~\ref{sec:apd1Xf}.
Given the homogeneous coordinates representation of all points on $\bar{P}$, the transformed curve in angle-uniform parallel coordinates can be generated.
For simplicity, we generate the transformed curve by transforming all samples on $\bar{P}$ and connecting these transformed points with short line segments.


\section{Visualization and User Interactions}
\label{sec:visualization}

\begin{figure}[t!]
	\centering    
	\includegraphics[width = \linewidth]{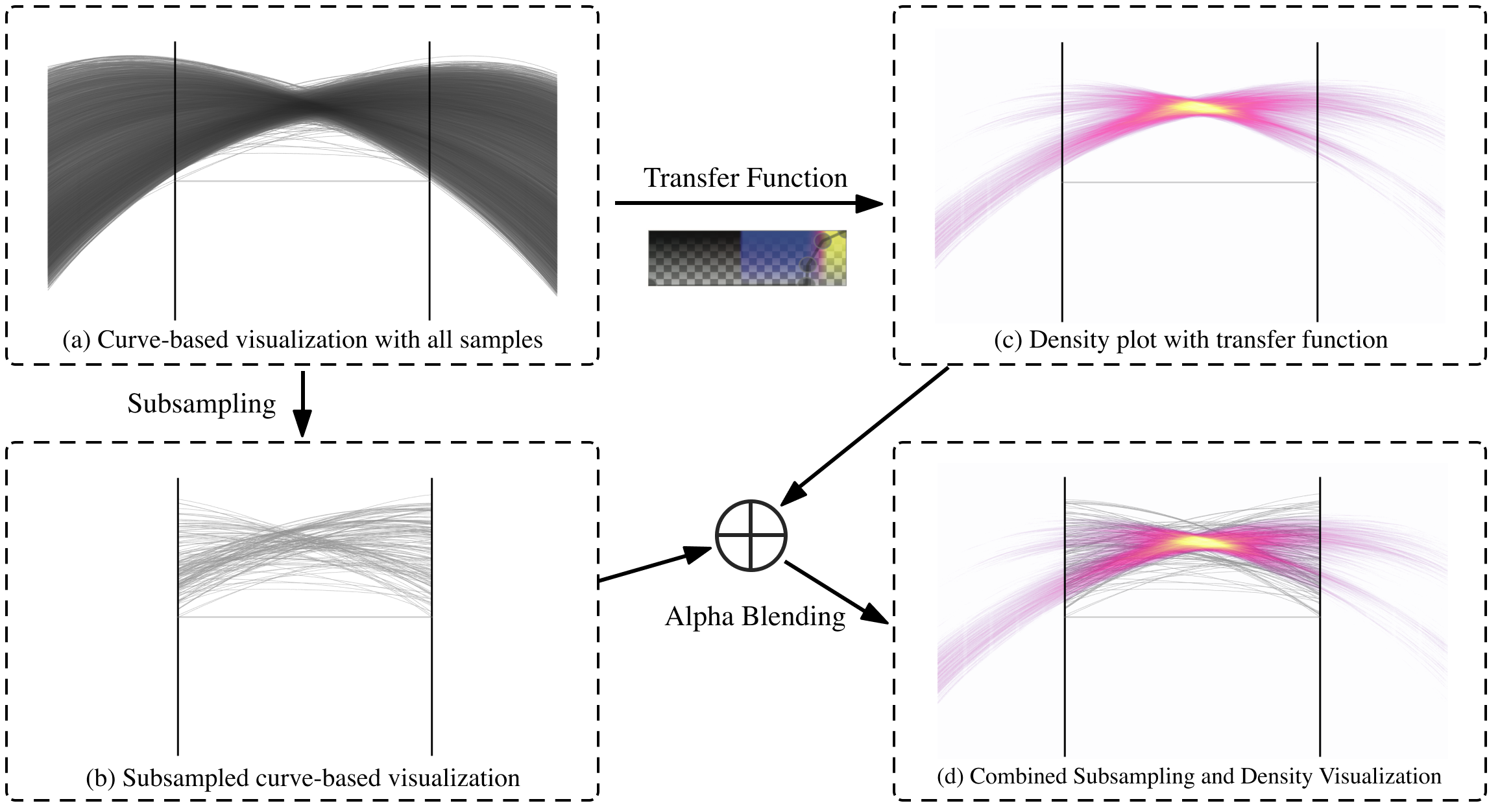}
	\caption{Workflow to generate a combined subsampling and density visualization.}
	\label{fig:docVis}
\end{figure}

Directly applying the transformation from Section~\ref{sec:method} to traditional parallel coordinates is sufficiently effective  to provide a symmetric visualization of negative and positive correlations as shown in Fig.~\ref{fig:docVis}(a).
However, if the dataset contains more than two attributes, the visualization is cluttered as curves of one pair of attributes extend to neighboring pairs, causing overdraw. 

In this section, we present a combined subsampling and density visualization approach that reduces visual clutter and facilitates pattern recognition in multidimensional data. \rev{Newly designed user interactions for exploratory data analysis, namely corner filtering and density plot brushing, are also explained.}

\subsection{Combined Subsampling and Density Visualization}

We propose a combined subsampling and density visualization approach for angle-uniform parallel coordinates. 
In this visualization, a ``curve'' layer (Fig.~\ref{fig:docVis}(b)) of transformed curves from the down-sampled dataset with a sampling rate of 5\%~\cite{Ellis:IV02} and a ``density'' layer of a density plot (see Fig.~\ref{fig:docVis}(c)) of the whole dataset are alpha-blended together.

The curve layer contains transformed curves only inside parallel coordinates as in traditional polyline-based parallel coordinates, i.e., only the part for $0\leq u \leq1$ is transformed; data entries of these curves are given by downsampling the multidimensional data in order to preserve discrete features of the data. 
Outliers in a dataset are important; a naive downsampling scheme could remove them and result in misleading visualizations. 
Therefore, we detect outliers in the multidimensional data domain~\cite{SugiyamaB2013_2} and mark the top $k$ outliers according to the distance ranking. 
These outliers are rendered in addition to the subsampled data entries in the visualization.

The density layer is a density representation of the full curves of the whole dataset. 
For simplicity, we use a binning strategy by accumulating curve density into an accumulation buffer.
Note that any density model can be used to generate the density plot. 
The accumulation buffer is rendered with a series of 1D transfer functions. 
Each pair of attributes has an associated 1D transfer function, comprising a colormap and an opacity map, to map a scalar value to a certain color and opacity. 

\begin{figure}[t]
	\centering
	\includegraphics[width =0.8\linewidth]{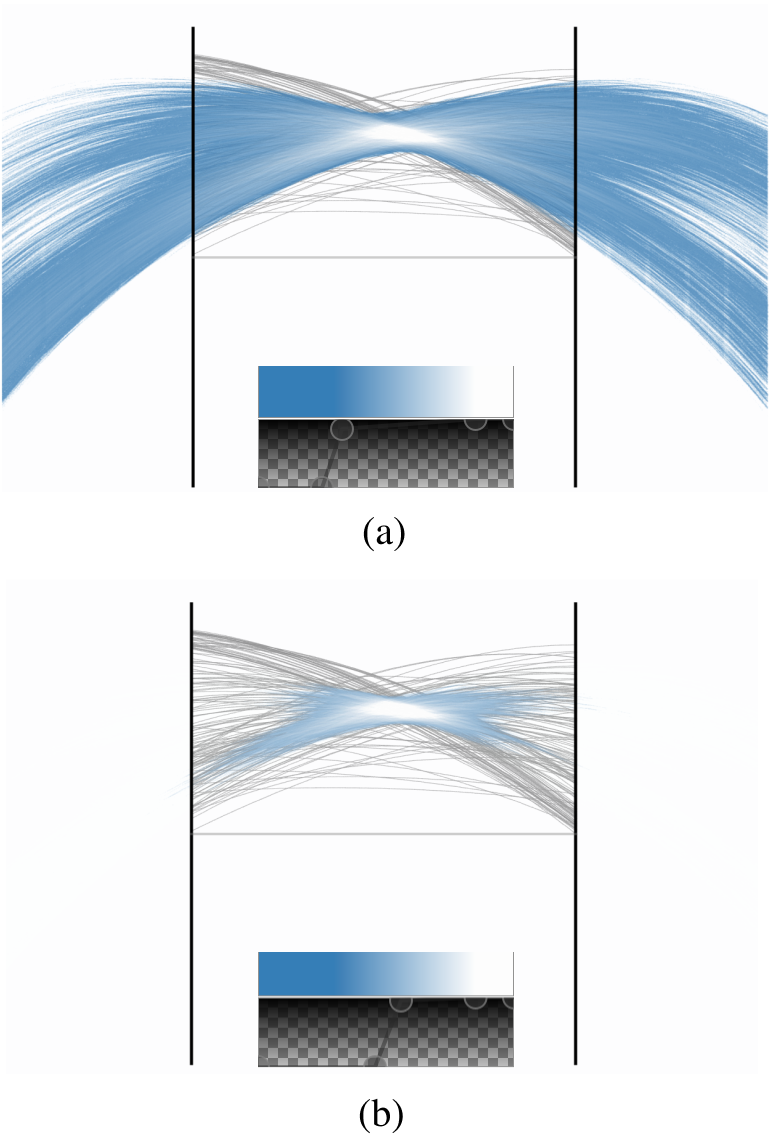}
	\caption{\rev{Combined subsampling and density visualizations with different transfer functions.}}
	\label{fig:tfcomp}
\end{figure}

\rev{
	Due to the high dynamic range of the density plot, non-linear opacity maps must be used. 
	Fig.~\ref{fig:tfcomp} shows how different transfer functions affect the rendering. A transfer function with a steep opacity ramp and a short high opacity range (see Fig.~\ref{fig:tfcomp}(b)) can remove the sparse regions that cover most of the screen area in Fig.~\ref{fig:tfcomp}(a) to highlight the region of high density. Therefore, we embed the transfer function adjustment interface into our interactive system so users can more flexibly generate the density layer.}

Finally, the combined subsampling and density visualization is generated by overlaying the density layer over the curve layer using alpha-blending (see Fig.~\ref{fig:docVis}). 
The visualization preserves both the local nature of discrete parallel coordinates and the global patterns from the density representation.
Compared to Fig.~\ref{fig:docVis}(a), the new visualization greatly reduces visual clutter.
High-density structures with both negative and positive correlations are clearly visible in Fig.~\ref{fig:docVis}(d).

\rev{
	In combined subsampling and density visualization, dense features are of interest because density is associated with the number of data entries that are linearly correlated. The spreading range of a feature is inversely associated with the strength of correlation of data: a perfectly correlated structure in data is shown as a single pixel of high density. The shape of a feature also reveals information: a shape stretched vertically indicates parallel linear structures in Cartesian space, while a horizontally spread shape indicates linear structures with smoothly changing orientations and similar intercepts. }

\subsection{Corner Filtering}
\rev{Adjusting the transfer function is a typical approach to feature filtering in density plots. However, opacity thresholds of transfer functions are based on the global data range without spatial information, which may not preserve important local features.}
\rev{For example, an intersection with a low density value is likely to be filtered out when the transfer function is set for intersections with high density values.}
Therefore, we apply corner detection, which uses a sliding window to check each local neighborhood and computes isotropy metrics for each pixel in the image, to further filter the density image.

\begin{figure}[t!]
	\centering
	\includegraphics[width = 0.9\linewidth]{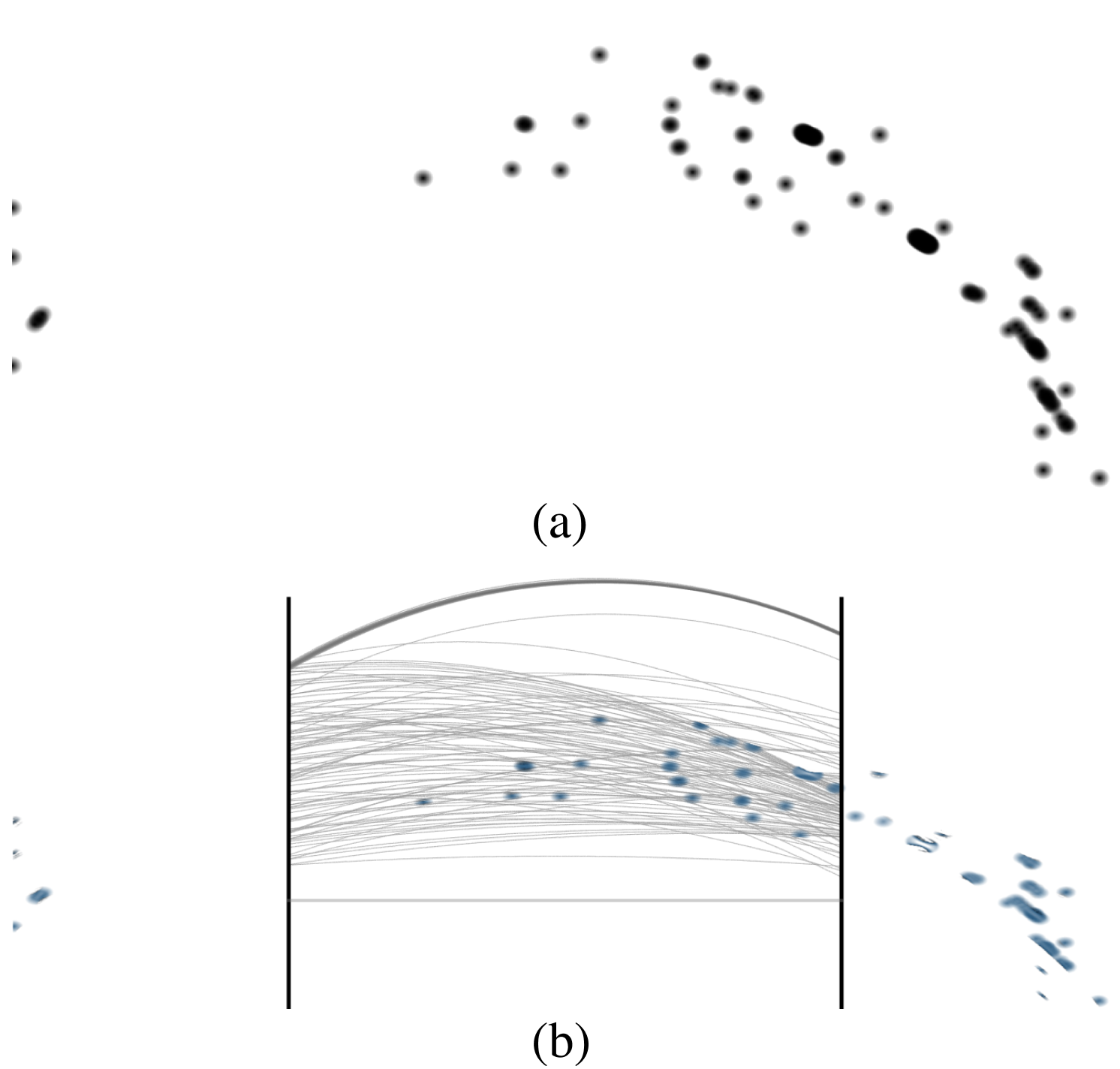}
	\caption{\rev{An image mask generated by corner filtering (a) and the overlaid result (b) that highlights the areas of curve intersections.}}
	\label{fig:mask}
\end{figure}

Specifically, the covariance matrix of data values of the neighborhood is calculated, and eigenvalues are computed and the pixel value of the metric image is assigned to the minimum eigenvalue.
Then, the metric image is normalized and used to build an image mask given a user-selected threshold during visualization.
In some cases, a local percentile filter that filters pixels below a certain percentile in the local neighborhood can be applied before corner filtering to keep corner pixels with low intensities.
The mask is created by drawing disks, with an opacity gradient that is fully opaque in the center and linearly falls off to fully transparent, into an accumulation buffer when the pixel value of the metric image is greater or equal to the threshold.
Fig.~\ref{fig:mask} shows an example of a mask generated by thresholding the corner metric image.

\subsection{Density Plot Brushing}
Brushing and linking are important in parallel coordinates; existing brushing methods for parallel coordinates select either ranges on axes~\cite{Ward:VIS:94,Fua:1999:Vis} or line segments in adjacent axes within a range of slopes~\cite{Hauser:VIS:2002}. 
A new type of brush is required for selecting regions in the density plot combining subsampling and density visualization of angle-uniform parallel coordinates (see Fig.~\ref{fig:synthetic}(d)). 

In essence, the brush tests all curves in a given 2D subspace of the data and highlights curves that pass through a user-drawn region.
In our current implementation, we redraw all entries of the data in the 2D image plane of angle-uniform parallel coordinates and test curves against the user-drawn region.
Both simple rectangle and arbitrary-shaped lasso brushes are supported.


\section{Case Studies}
\rev{In this section,} we demonstrate the advantages of combined subsampling and density visualization of angle-uniform parallel coordinates over traditional parallel coordinates \rev{and their variants, as well as} 2D scatterplots, through synthetic and real-world examples.

\subsection{Synthetic Dataset} \label{sect:synthetic_data}
\rev{To help readers better understand angle-uniform parallel coordinates as a new visualization technique, we constructed a synthetic dataset that contains several linear correlation structures. The visualization results using this synthetic dataset highlight the benefits of angle-uniform parallel coordinates in analyzing linear relationships. This example also demonstrates that visual pattern analysis can be aided by brushing and corner filtering.}

\rev{The synthetic dataset was constructed by sampling a set of 2D Gaussian distributions with covariance matrices of corresponding angles. Specifically, the data points were sampled from twelve line segments, including three lines of $-5\degree$, $15\degree$, and $50\degree$, respectively, and nine parallel lines of $30\degree$. The $15\degree$ line, among others, was most densely sampled.}

\begin{figure*}[ht]
	\centering
	\includegraphics[width = \linewidth]{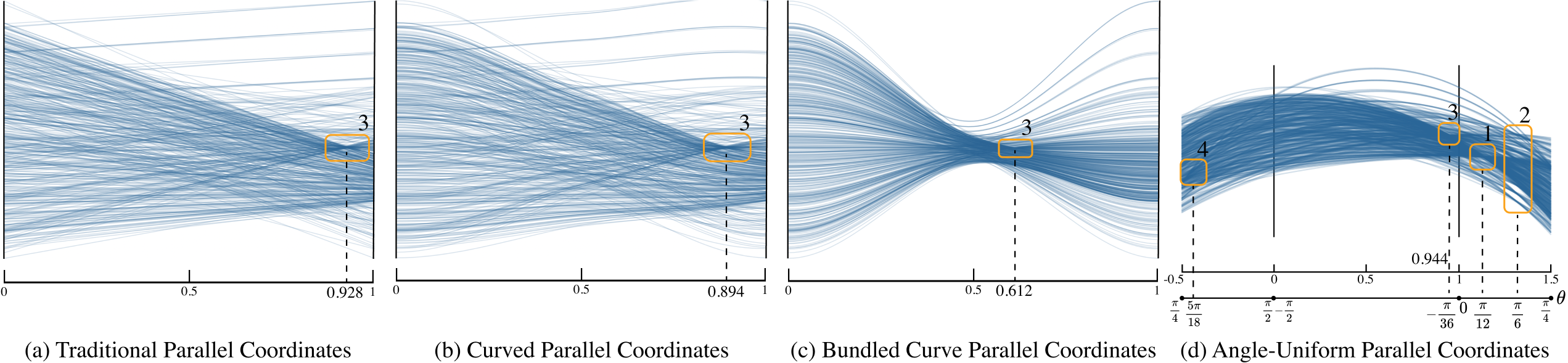}
	\caption{\rev{Comparison of four variants of parallel coordinates using the synthetic dataset (see Section~\ref{sect:synthetic_data}). Line intersections representing strong positive and negative linear correlations are marked with orange boxes and numbers (in agreement with Fig.~\ref{fig:synthetic}). The horizontal coordinates of intersections are marked and measured.}}
	\label{fig:synthetic_compare}
\end{figure*}

\subsubsection{Visual Pattern Interpretation}

Fig.~\ref{fig:synthetic}(a) shows that only one strong crossing pattern of negative correlation (highlighted by the orange box) can be seen when using traditional parallel coordinates. 
The discrete nature of the data can be partly visualized by clusters of line segments. 
With combined subsampling and density visualization of angle-uniform parallel coordinates (see Fig.~\ref{fig:synthetic}(c)), and a transfer function with a blue-to-white color map, much more insight is given; several visual patterns of high density (in light blue or white) can be seen.

Given these visual patterns, we are able to build a mental image of the data (labeled in Fig.~\ref{fig:synthetic}(c) and Fig.~\ref{fig:synthetic_compare}(d)): 
\begin{enumerate}[noitemsep, topsep=3pt]
\item A high-density line structure with an orientation of roughly ${\pi}/{12}$ ($15\degree$).
\item  Several parallel line structures oriented toward ${\pi}/{6}$ ($30\degree$).
\item  The same negatively oriented linear structure seen in the traditional parallel coordinates still occurs.
\item  A medium-density line structure with an angle a little greater than ${\pi}/{4}$ (around $50\degree$).
\end{enumerate}

A clearer visualization (Fig.~\ref{fig:synthetic}(d)) is achieved by filtering out weak visual patterns using corner filtering and brushing on strong visual patterns with lassos.
With this diagram, we are able to validate our interpretation of the dataset as seen in the scatterplot in Fig.~\ref{fig:synthetic}(d).
We associate the brushed patterns in parallel coordinates with the corresponding linear structures in the Cartesian data domain by highlighting the curves with the same colors as the brushes.

\begin{figure*}[!tb]
	\centering
	\includegraphics[width=\linewidth]{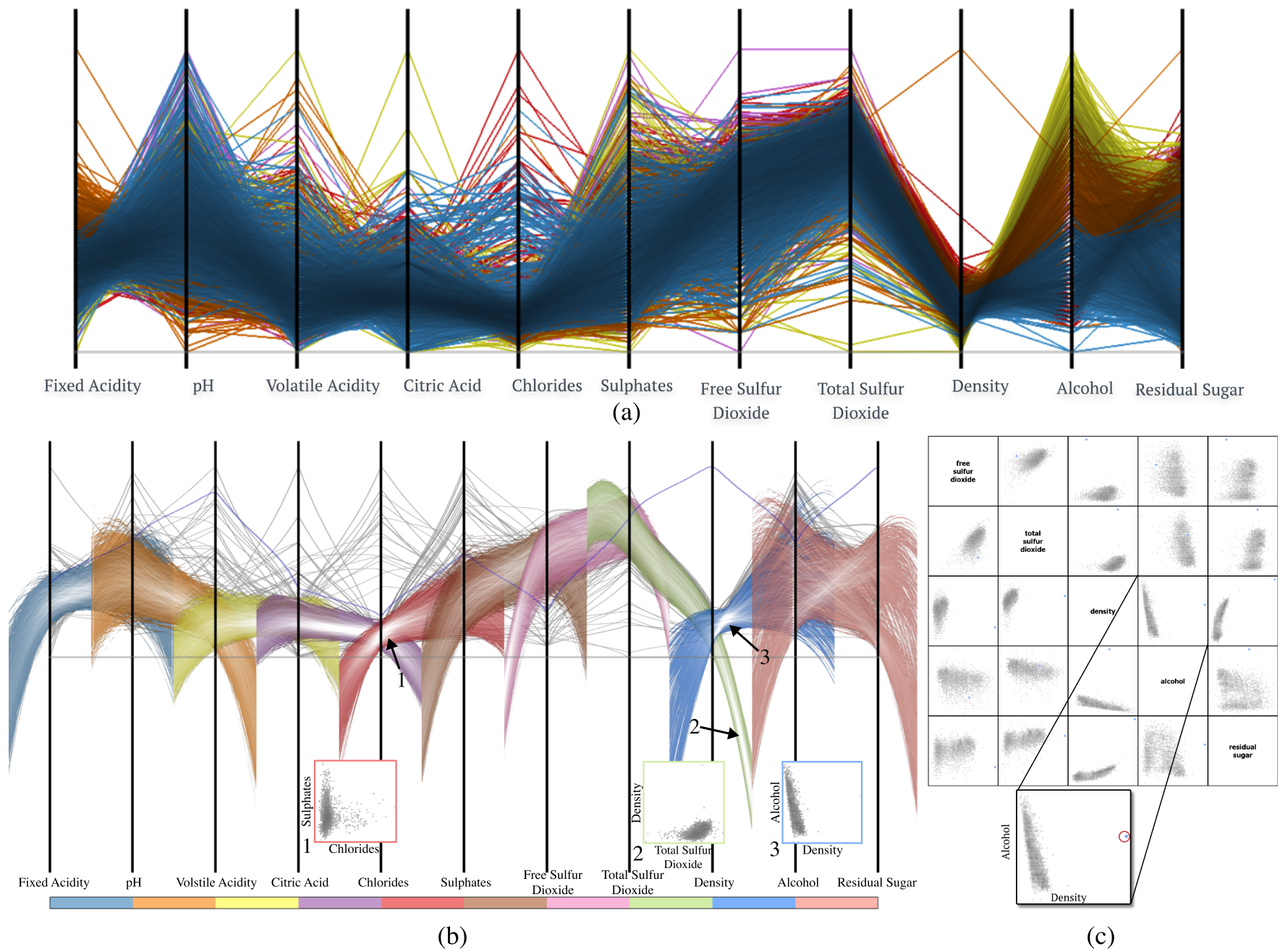}
	\caption{Traditional parallel coordinates colored by $K$-means clustering (a), the combined subsampling and density visualization of angle-uniform parallel coordinates (b), and a 5D part of the SPLOM (c) of the white wine quality dataset. Concentrated patterns are marked with numbers and their associated 2D histograms are shown. \rev{Our} visualization reduces clutter and preserves both global and local correlation patterns. An outlier of the \emph{Density} attribute is marked \rev{with a blue curve} in (b) and \rev{blue points} in (c).}
	\label{fig:teaser}
\end{figure*}

\subsubsection{\rev{Comparison to Variants of Parallel Coordinates}}

\rev{
	To further demonstrate the benefits of angle-uniform parallel coordinates (AUPC), we compared AUPC, traditional parallel coordinates (PC), and  variants of the latter. 
	Since AUPC deforms the image plane of parallel coordinates, and thereby transforms polylines into curves, we chose two representative curve variants of parallel coordinates:  curved parallel coordinates (CPC), and bundled curve parallel coordinates (BCPC)~\cite{Heinrich:2012:EBT}, which achieve better visual separation between data clusters and provide a clearer overview of the dataset.}

\rev{
	We downsampled the synthetic dataset at a rate of 4\% to avoid cluttering the image planes and ensure that lines and curves could be traced.
	For CPC and BCPC, we used an open-source Javascript implementation~\cite{PCgithub}, setting the smoothness scale $\alpha=1/6$ and bundling strength $\beta=0$ for CPC, and $\alpha=1/6$, $\beta=0.8$ for BCPC as the recommended values for best balancing correlation detection and cluster visualization~\cite{Heinrich:2012:EBT}. 
	For AUPC, we did not apply the combined subsampling and density visualization to ensure fairness of comparison.}

\rev{
	As Fig~\ref{fig:synthetic_compare} shows, all methods can reveal the negative correlation pattern (number 3), but only AUPC reveals positive correlations.
	As the horizontal coordinate of PC, CPC, and BCPC is not linearly mapped to the angle of Cartesian lines, the slope of the disclosed negative linear structure can easily be  misjudged, especially in BCPC, where the intersection point is moved significantly toward the middle.
	Although individual positive correlations can be identified by expert users through a small bundle of disjoint curves in BCPC, only a vague impression of a large proportion of positive correlations can be obtained from Fig.~\ref{fig:synthetic_compare}(c), where multiple positive and negative correlations are involved and occlude each other.}

\rev{In contrast, our AUPC can clearly show both negative and positive correlations, and users can quantify the slopes of Cartesian lines accurately from the horizontal coordinates of the intersections.}

\begin{figure*}[t!]
	\includegraphics[width = \linewidth]{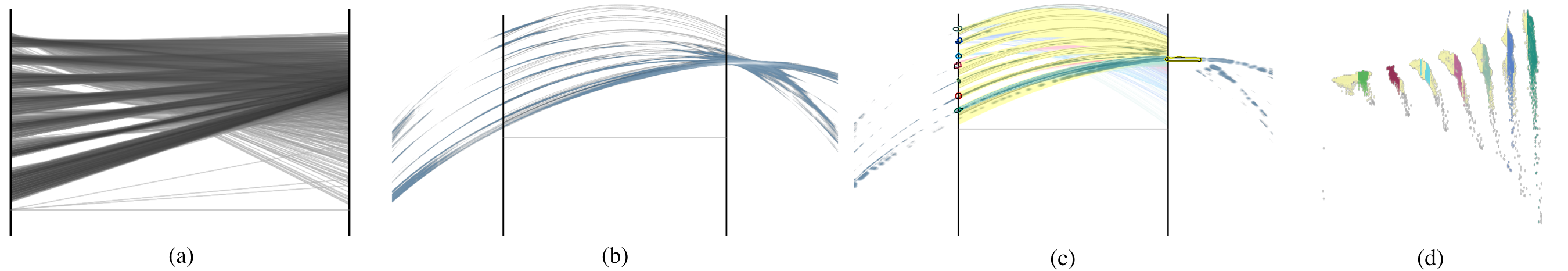}
	\caption{Visualization of attributes \emph{pressure} and \emph{temperature} of a downsampled version of the Hurricane Isabel simulation using (a)~traditional parallel coordinates, (b)~combined subsampling and density visualization of angle-uniform parallel coordinates, (c)~corner filtering and density plot brushing, and (d)~highlighted patterns in the data domain.}
	\label{fig:isabel2D}
\end{figure*}

\subsection{White Wine Quality}

The white wine quality data~\cite{Cortez:2009:MWP:1628313.1628356} contains 4897 samples of 11 chemical properties as continuous variables and a subjective quality attribute as an ordinal variable.
We discard the quality attribute and rearrange the order of axes to generate the visualization in Fig.~\ref{fig:teaser}.


\rev{Traditional parallel coordinates have no color encoding for unlabeled data. However, in combined subsampling and density visualization, the color map can be used to discriminate different attribute pairs when drawing the density plot.
	For a fair comparison of our method to traditional parallel coordinates in terms of perception, we add the color channel to traditional parallel coordinates by first clustering the data points using the $K$-means algorithm and then coloring them according to cluster labels.}
With traditional parallel coordinates~(see Fig.~\ref{fig:teaser}(a)), it is difficult to visualize correlations of attributes and the coloring is not very effective due to occlusion.

In contrast, the combined subsampling and density visualization of angle-uniform parallel coordinates (see Fig.~\ref{fig:teaser}(b)) allows a better understanding of correlations and visualization of outliers. 
Note that this plot employs the optional vertical scaling function.
Strong correlations are shown as concentrated high-density patterns for several attribute pairs: \emph{chlorides--sulphates} (marked as number~1), \emph{total sulfur dioxide--density} (number~2), and \emph{density--alcohol} (number~3).
It is clear from these density patterns that:
\begin{enumerate}[noitemsep, topsep=3pt]
\item the regression line of \emph{chlorides--sulphates} is almost vertical (see also histogram~1);
\item the slope of the regression line of \emph{total sulfur dioxide--density} is between $0\degree$ and $45\degree$ (see histogram~2);
\item \emph{density--alcohol} has a negative correlation with the angle of the regression line close to $-{3\pi}/{8}$ ($-67.5\degree$, see histogram~3).
\end{enumerate}

This pair-wise correlation information can also be visualized  in a SPLOM (see Fig.~\ref{fig:teaser}(c)) that shows a 5D subspace of the data.
However, following data across higher dimensions (e.g., $\geq$ 8D) is much easier with parallel coordinates and angle-uniform parallel coordinates.  
For example, an outlier in the \emph{density} attribute \rev{marked by a blue curve} 
in the combined subsampling and density visualization of angle-uniform parallel coordinates can be easily traced across all dimensions.
It is already challenging to locate attribute values of outliers in the 5D SPLOM, let alone trace all attribute values in the full SPLOM.

Another advantage of parallel coordinates and angle-uniform parallel coordinates over SPLOMs is that they scale better with increasing dimensionality.
The limited display area causes the outliers to be rendered too small to be seen in the SPLOM.
Therefore, repetitive zooming-and-panning (see Fig.~\ref{fig:teaser}(c) inset) is required for users to access detailed information. 
This excessive context switching can easily break the user's mental map.

\subsection{Hurricane Isabel}
The Hurricane Isabel simulation dataset is a widely used multivariate volume dataset containing $500\times 500 \times 100$ spatial samples~\cite{HurricaneIsabel:2004:VisContest}. 
Here, we use attributes of time step 25.

\subsubsection{Analysis of Two Attributes}

We analyze the \emph{pressure} and \emph{temperature} attributes with a downsampled dataset of resolution $50\times 50\times 10$.
The traditional parallel coordinates plot is shown in Fig.~\ref{fig:isabel2D}(a).
Some basic discoveries can be made from this visualization, e.g., several discrete line groups exist, indicating the down-sampled nature of the data, and a large portion of samples have low pressure. 
However, we identify drawbacks at two scales: at the global scale, it is impossible to learn about the linear relationships of the whole dataset; 
at a finer scale, it is unclear how samples inside one line group correlate or whether they have linear relationships. 
With combined subsampling and density visualization of angle-uniform parallel coordinates (see Fig.~\ref{fig:isabel2D}(b)), one can easily read off information about linear relationships at different scales.
Both local and global patterns can be highlighted in the data domain (Fig.~\ref{fig:isabel2D}(d))  by brushing on strong patterns after corner filtering (Fig.~\ref{fig:isabel2D}(c)).

\subsubsection{Multiple Attributes}
In the above examples, we have seen that angle-uniform parallel coordinates can convey more information than traditional parallel coordinates.
A major benefit of parallel coordinates and its variants is that it is easier to visualize multidimensional relationships of specific samples compared with SPLOM. 
However, with data that contains only two attributes, they do not outperform scatterplots in terms of linear relationship analysis.

Parallel coordinates suffer from visual clutter as data size increase. 
Our combined subsampling and density visualization approach integrates the benefits of uncluttered curves for multidimensional attribute tracking and density plots for linear relationship analysis. 
To demonstrate the advantages of combined subsampling and density visualization of angle-uniform parallel coordinates, we conducted a further experiment using four attributes of the Isabel dataset: \emph{speed, height, pressure} and \emph{temperature}. 
Note that the dataset was Monte-Carlo sampled with a sampling rate of 1\%.

\begin{figure*}[t!]
	\centering
	\includegraphics*[width =\linewidth]{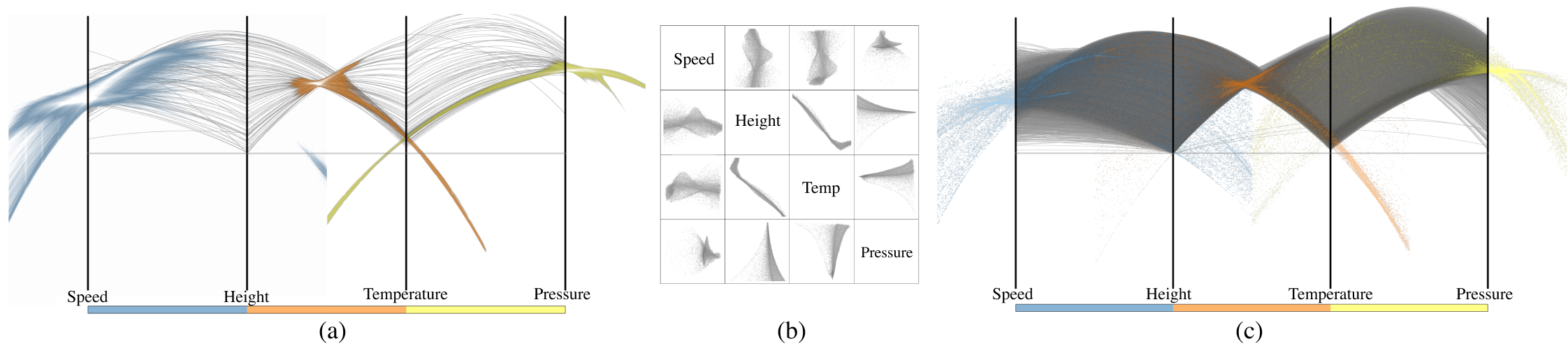}    
	\caption{Four attributes of the Hurricane Isabel data visualized using (a) combined subsampling and density visualization of angle-uniform parallel coordinates, (b) SPLOM, and (c) indexed-point visualization of angle-uniform parallel coordinates.
	}
	\label{fig:isabel3D}
\end{figure*}

By examining the density plot and sub-sampled curves, it is clear from Fig.~\ref{fig:isabel3D}(a) that:
\begin{enumerate}[noitemsep, topsep=3pt]
\item the lower half of \emph{height} has a strong negative correlation with \emph{temperature}, and the Cartesian angle is around $-45\degree$ \item these \emph{temperature} samples have a strong positive correlation with \emph{pressure} with a small positive angle
\item the top part of \emph{height} has lowest \emph{temperature} and medium \emph{pressure} 
\item one outlier has lowest \emph{height}, highest \emph{temperature} and lowest \emph{pressure}.
\end{enumerate}

With a SPLOM (see Fig.~\ref{fig:isabel3D}(b)), linear information can be visualized. However, it is impossible to learn the relationship between attributes of specific samples without brushing and linking. 

Comparing the combined subsampling and density visualization of angle-uniform parallel coordinates (see Fig.~\ref{fig:isabel3D}(a)) with the angle-uniform extension of 1-flat indexed-points~\cite{Zhou:TVCG:inprint} (Fig.~\ref{fig:isabel3D}(c)), it is noticeable that:
\begin{enumerate}[noitemsep, topsep=3pt]
\item indexed points of local correlations form similar global patterns as for the density plot
\item indexed points reveal more details inside  high density regions
\item global density patterns cover regions that are not seen in the indexed-point visualization, e.g., the thin yellow band that stretches across the last sub-dimension. 
\end{enumerate}


\section{Discussion}
\label{sec:discussion}

\subsection{\rev{Representation Ability and Intuitiveness}}\rev{
We now consider the trade-off between representation ability and intuitiveness.	Although regions II and III (see Fig.~\ref{fig:slopenOrientation}(b)) intrinsically exist in parallel coordinates as given by the point--line duality, they are discarded in almost every use case because of the problems caused by the infinite image plane, to make parallel coordinates easy to interpret.
	However, this simplification prevents traditional parallel coordinates from representing positive correlations, which is a limitation in visual analysis tasks.
	In our angle-uniform parallel coordinates, we compress the image plane from being infinite to finite size through geometric deformation, allowing regions II and III to be fully visualized.
	Although our method sacrifices intuitiveness compared with other variants of parallel coordinates, it restores the full representation ability of the original parallel coordinates model and extends its application by allowing quantitative visual interpretation of Cartesian lines.
}

\subsection{\rev{Information Loss}}\rev{
	Our method of combined subsampling and density visualization is designed so as to minimizing information loss during the visualization process that blends two image layers.
	In the curve layer, local patterns like outliers are detected in the data domain and preserved during subsampling. In the density layer, the density plot is computed from the whole dataset and therefore contains global information.
	Combining these two techniques can preserve important information from the input dataset while significantly reducing visual clutter. 
	One limitation, however, is that users may lose track of some curves covered by high-density areas. This problem can be addressed by adjusting the transfer function and opacity values in our interactive system.
}

\subsection{\rev{Future Work}}\rev{
	We plan to improve our method in three directions in future.
	Firstly, only 1-flat indexed points are supported by angle-uniform parallel coordinates, so a new mathematical model for angle-uniform representations of higher-dimensional $p$-flats~\cite{Zhou:TVCG:inprint} is needed.
	Secondly, a full-fledged visual analysis system should be built with machine learning methods to help the user cluster the density plot with ease.
	Lastly, we intend to conduct a comprehensive user study to evaluate the utility of various parallel coordinates methods including angle-uniform parallel coordinates for different tasks. 
}


\section{Conclusions}
\label{sec:conclusion}
In this paper, we have proposed angle-uniform parallel coordinates, a general data-independent parallel coordinate model that deforms traditional parallel coordinates to provide more insight without destroying important geometric relationships.
The biggest benefit of our method is that a symmetric representation of Cartesian lines of any orientation is achieved.
Our combined subsampling and density visualization approach allows effective visual analysis of linear correlations, and regressions of multidimensional datasets can be performed in angle-uniform parallel coordinates without context switches. 
The data independence of our method makes it applicable to all multidimensional datasets.
Further, the method can be easily implemented and incorporated within existing parallel coordinates frameworks.


\CvmAck{
LZ acknowledges support from the Data for Better Health Project of Peking University-Master Kong, YW from the National Science Foundation of China (62132017), and DW from the Deutsche Forschungsgemeinschaft (DFG) Project-ID 251654672-TRR 161.
}

\bibliographystyle{CVM}

{\normalsize  \bibliography{ref}}

\Author{Photo_kaiyi}{Kaiyi Zhang}{
is a third-year master's student in the School of Computer Science and Technology, Shandong University. His research interests include text visualization and visual analysis. }

\Author{Photo_liang}{Liang Zhou}
{is an assistant professor at the National Institute of Health Data Science, Peking University. His research interests include scientific and information visualization, visual perception, and visual analytics for health science.}

\Author{Photo_chenlu}{Lu Chen}{
is currently a Ph.D. student at the State Key Lab of CAD\&CG, Zhejiang University. He obtained his B.Eng. degree in computer science from Shandong University in 2022. His research interests lie primarily in 3D computer vision and information visualization.}

\Author{Photo_shitong}{Shitong He}{
is a senior student at Taishan College, Shandong University. His research interests include scientific visualization and information visualization.}

\Author{weiskopf}{Daniel Weiskopf}
{is a professor at the Visualization Research Center (VISUS) of the University of Stuttgart, Germany. His research interests include visualization, visual analytics, eye tracking, GPU methods, computer graphics, human-computer interaction, augmented and virtual reality, and special and general relativity.}

\Author{Photo_yunhai}{Yunhai Wang}
{is a professor in the School of Computer Science and Technology at Shandong University. He serves as an associate editor of Computer Graphics Forum. His interests include scientific visualization, information visualization and computer graphics.}

\appendices{
\section{Appendix I Transformation of 1-flat Indexed Points}

We detail the derivation of special cases of $v$ and the scaling function $s(u)$ in the transformations from traditional to angle-uniform parallel coordinates. 

\subsection{Special cases of $v$ in Equation~\ref{eqn:fullv}}
\label{sec:apdvertXf}

We address the exceptions using the limit method, i.e., taking the vertical coordinate of a transformed point $\bar{l}_t = (u_t,v_t)$ whose horizontal coordinate is infinitely close to the point of question $\bar{l}_c = (u_c,v_c)$ as the vertical coordinate of $\bar{l}_c$:
\begin{equation}
v_{c} = v_{t}\;, \\
\text{where\space} |u_t-u_c| = \delta\;, \delta > 0, \delta \to 0, \delta \in\mathbb{R}\;.
\nonumber
\end{equation}
For the case of $u$ at $u_{0.5}=0.5$, we approach the point $(u_{0.5},v_{0.5})$ by averaging the vertical coordinates of points $(0.5-\delta,v_{t_{0.5^{-}}})$, $(0.5+\delta, v_{t_{0.5^{+}}})$ with $\delta\to 0$ from the left and the right:
\begin{align}
&v_{0.5} = 0.5\cdot(v_{t_{0.5^{-}}} + v_{t_{0.5^{+}}})\;,\\
&\text{where\space} u_{t_{0.5^{-}}}=0.5 - \delta\;,u_{t_{0.5^{+}}}=0.5 + \delta\;.\nonumber
\nonumber
\end{align}
If $u=-0.5$, the point is approached from the right; for the case of $u=1.5$, the point is approached from the left:
\begin{equation} \small
\begin{cases}
v_{-0.5} = \displaystyle\lim_{\delta\to0}v_{-0.5+\delta} =v_{t_{-0.5}}\;,&\text{where\space} u_{t_{-0.5}}=-0.5+\delta\\
v_{1.5} = \displaystyle\lim_{\delta\to0}v_{1.5-\delta} = v_{t_{1.5}}\;,&\text{where\space} u_{t_{1.5}}=1.5-\delta\;.
\nonumber
\end{cases}
\end{equation}
We then get all cases for the transformed vertical coordinate as in Equation~\ref{eqn:fullv}.

\subsection{Derivation of $s(u)$}
\label{sec:apdsu}

To ensure the smoothness of the scaling function, we use a cubic spline: $s(u)$.  
In order to satisfy goals~\ref{itm:subspaceC},~\ref{itm:crossspaceC},~\ref{itm:symmetric} and to avoid oscillations of transformed curves that hinder the perception, we specify eleven control points that the cubic spline $s(u)$ must pass through: 
\begin{itemize}[noitemsep,topsep=3pt]
	\item $\bar{l}_{C}^1$ at $u=-0.5$: $(-0.5, 1.306)$.
	\item $\bar{l}_{C}^2$ at $u=-0.25$: $(-0.25, 1.153)$.
	\item $\bar{l}_{C}^3$ at $u=0$: $(0, 1)$.
	\item $\bar{l}_{C}^4$ at $u=0.1$: $(0.1, 0.9312)$.
	\item $\bar{l}_{C}^5$ at $u=0.25$: $(0.25, 0.8555)$.
	\item $\bar{l}_{C}^6$ at $u=0.5$: $(0.5, 0.812)$.
	\item $\bar{l}_{C}^7$ at $u=0.75$: $(0.75, 0.8555)$.
	\item $\bar{l}_{C}^8$ at $u=0.9$: $(0.9, 0.9312)$.
	\item $\bar{l}_{C}^9$ at $u=1$: $(1, 1)$.
	\item $\bar{l}_{C}^{10}$ at $u=1.25$: $(1.25, 1.153)$.
	\item $\bar{l}_{C}^{11}$ at $u=1.5$: $(1.5, 1.306)$.
\end{itemize}
Since the spline is cubic, the scaling function $s(u)$ has $C^2$ continuity. 
The fitted spline with the scaling data and control points are illustrated in Figure~\ref{fig:spline}.
\begin{figure}[!htb]
	\centering
	\includegraphics[width=0.7\linewidth]{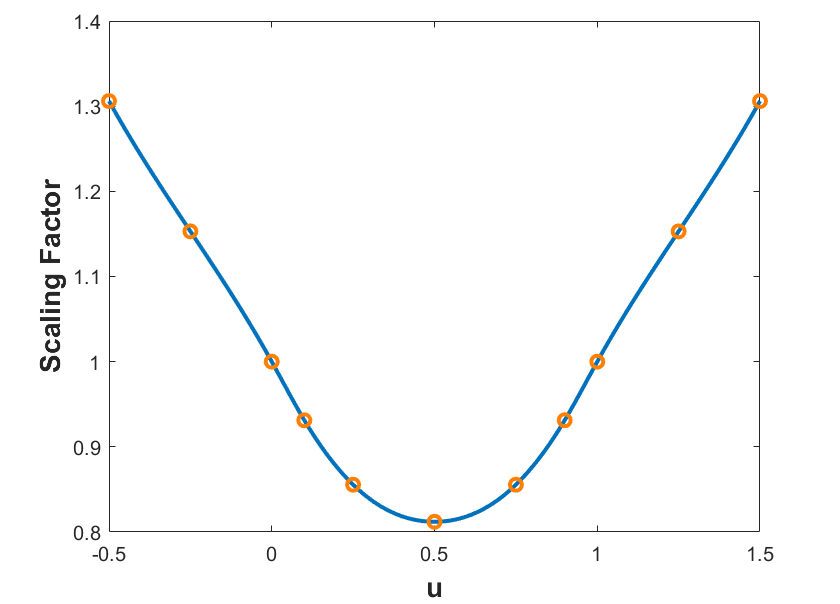}
	\caption{The scaling function $s(u)$ as a cubic spline. The control points $\bar{l}_C$ of the cubic spline are marked as orange circles.}
	\label{fig:spline}
\end{figure}

\section{Transformation of Parallel Coordinates Lines}
\label{sec:apd1Xf}

The line equation of a parallel coordinates line $\bar{P}$ of a 2D Cartesian point $P = (p_1,p_2)$ can be derived by the point--line duality:
\begin{equation}
\bar{P}: y = (p_2-p_1)x + p_1\;.
\end{equation}
Therefore, for $\theta \neq \frac{\pi}{4}$, the 1-flat indexed point on $\bar{P}$ with $\theta$, $\bar{l}_{\bar{P}}^{\theta} = (x_{\bar{P}}^{\theta},y_{\bar{P}}^{\theta})$, can be calculated as:
\begin{equation}
\begin{cases}
x_{\bar{P}}^{\theta} &= \frac{1}{1-\tan(\theta)}\;,\\
y_{\bar{P}}^{\theta} &= (p_2 - p_1)x_{\bar{P}}^{\theta}+p_1\;.
\end{cases}
\end{equation}
Further, with Equation~\ref{eqn:lineInPCP}, we are able to compute the slope and intercept of the associated line of $\bar{l}_{\bar{P}}^{\theta}$ in Cartesian coordinates:
\begin{equation}
\begin{cases}
a_{\bar{P}}^{\theta} &= \tan(\theta)\,\\
b_{\bar{P}}^{\theta} &= \frac{y_{\bar{P}}^{\theta}}{x_{\bar{P}}^{\theta}}\;.
\end{cases}
\end{equation}
For the case of $\theta = \frac{\pi}{4}$, the transformed 1-flat indexed point is calculated by the limit approach discussed in Section~\ref{sec:vertXf} and Appendix~\ref{sec:apdvertXf}, i.e., we choose a $\theta$ that is infinitely close to $\frac{\pi}{4}$.
If the sample point is close to
$-\infty$, $\theta = \frac{\pi}{4} + \Delta$, and if the point is close to $+\infty$, $\theta = \frac{\pi}{4} - \Delta$, where $\Delta > 0, \Delta \to 0$.
When $\theta = \pm\frac{\pi}{2}$, because of the symmetry of angle-uniform parallel coordinates, the point location is intact after the transformation, i.e., on the left axis ($u=0$) at location $v=y$.
Therefore, we obtain $[c_1,c_2,c_3]$ for all cases:
\begin{equation}
\begin{cases}
c_1 = -a_{\bar{P}}^{\theta},\;c_2 = 1,\;c_3=-b_{\bar{P}}^{\theta},\;\theta\neq\pm\frac{\pi}{2}, \theta\neq\frac{\pi}{4}\\
c_1 = 1,\;c_2=0,\;c_3=-p_1\;,\quad\quad\quad\quad\;\theta=\pm\frac{\pi}{2}\\
c_1 = -a_{\bar{P}}^{\frac{\pi}{4} \pm \Delta},\;c_2=-1,\;c_3=-b_{\bar{P}}^{\frac{\pi}{4} \pm \Delta}\;,\theta = \frac{\pi}{4}\;.\\
\end{cases}
\end{equation}}
\end{document}